Short Paper

# Barriers and Challenges of Computing Students in an Online Learning Environment: Insights from One Private University in the Philippines

Bernie S. Fabito
Center for Innovation and Entrepreneurship, National University-Manila
bsfabito@national-u.edu.ph
(corresponding author)

Arlene O. Trillanes
College of Computing and Information Technologies, National University-Manila
aotrillanes@national-u.edu.ph

Jeshnile R. Sarmiento
College of Computing and Information Technologies, National University-Manila
sarmientojr@students.national-u.edu.ph

## Abstract

*Purpose* – While the literature presents various advantages of using blended learning, policymakers must identify the barriers and challenges faced by students that may cripple their online learning experience. Understanding these barriers can help academic institutions craft policies to advance and improve the students' online learning experience. This study was conducted to determine the challenges of computing students in one private University in the Philippines during the period where the entire Luzon region was placed under the Enhanced Community Quarantine (ECQ) as a response to the COVID-19 pandemic.

*Method* – A survey through MS Forms Pro was performed to identify the experiences of students in online learning. The survey ran from March 16 to March 18, 2020, which yielded a total of 300 responses.

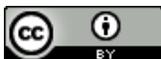

*Results* – Descriptive statistics revealed that the top three barriers and challenges encountered by students were 1.) the difficulty of clarifying topics or discussions with the professors, 2.) the lack of study or working area for doing online activities, and 3.) the lack of a good Internet connection for participating in online activities.

*Conclusion* – It can be concluded that both students and faculty members were not fully prepared to undergo full online learning. More so, some faculty members may have failed to adapt to the needs of the students in an online learning environment.

*Recommendations* – While the primary data of the study mainly came from the students, it would also be an excellent addition to understand the perspective of the faculty members in terms of their experiences with their students. Their insights could help validate the responses in the survey and provide other barriers that may not have been included in the study.

*Practical Implications* – The study has revealed gaps in online learning as experienced by the computing students, which can help the management craft a holistic framework encompassing both the learners and teachers for a successful online learning experience for the students.

## INTRODUCTION

The rapid advancement of technology has dramatically revolutionized our education system. The use of technology (e.g., Internet, Mobile Devices, Augment Reality, Virtual Reality) has led to pedagogical evolution (Chilton & McCracken, 2017), creating a paradigm shift in education. In recent years, studies have been conducted to determine how Mobile Learning, which makes use of mobile devices, create a better learning experience for learners (Almaiah, Jalil, & Man, 2016; Alrasheedi, Capretz, & Raza, 2016; Fabito, 2017). As more mobile devices become connected to the Internet, the growing implication for online learning adoption has been increasing.

Online learning is one of the methods used by Higher Educational Institutions (HEIs) to support traditional learning (Janse van Rensburg, 2018). The use of modern tools such as email, audio, and video teleconferencing for online learning (Schindler, Burkholder, Morad, & Marsh, 2017) are some of the tools that can be used by academicians (Rabacal, 2018) to augment the delivery of education. The combination of both using online learning and face to face instruction has paved the way for what is now known as Blended Learning (Rasheed, Kamsin, & Abdullah, 2020). Blended learning is advantageous among faculty members and learners due to the limitations brought about by traditional learning. Students who are restricted to physical space and time in a classroom setting may gain access to online resources (Calamlam, 2016), collaborate among students, and connect with faculty members in real-time through PCs, Laptops, Mobile Phones, and Tablets.

Although the literature has presented the benefits of Blended Learning in the Philippine setting (Calamlam, 2016; Rabacal, 2018), it is crucial to understand the barriers that hamper students to embrace online learning. In times of calamities and natural disasters that force HEIs to shift to online learning suddenly, institutions must understand the accessibility of students to participate in online learning before it happens. The success of online learning can occur when the management can address the issues of students in the conduct of online learning.

In the field of computing, the requirement varies with other programs as some professional subjects would require devices with specific hardware needs for programming and other computing courses ("Student Computing Requirements - Information Technology - University of Florida," n.d.). Understanding this crucial component is a must to implement online learning without relaxing the needed course outcomes effectively.

With this said, the purpose of the study was to determine the challenges and barriers of computing students in online learning, particularly in one private University in the Philippines. While there seemed to be a plethora of literature that sought to understand possible barriers and challenges of blended learning (Janse van Rensburg, 2018; Rasheed et al., 2020; Sun & Chen, 2016), what made this study unique was its general focus on computing students as the needed requirement for blended learning varies with other degree programs. One other example where specific requirements vary in online education is the nursing degree program (Smith, Passmore, & Faught, 2009).

## METHODOLOGY

This study was conducted when Luzon, one of the major islands in the Philippines, was placed under Enhanced Community Quarantine (ECQ), which started on March 17, 2020, until April 13, 2020, in response to the COVID-19 pandemic ("Community Quarantine over the Entire Luzon," n.d.). During this period, classes all over Luzon were suspended, including Metro-Manila. However, as early as March 9, 2020, classes in Metro-Manila were already suspended due to the growing cases of COVID-19. This prompted most HEIs to immediately shift to online learning. Students and faculty members were barred from entering the school campus and were asked to use online learning so as not to disrupt the academic calendar. Starting on March 9, most schools went online.

From the period of March 9 to March 16, students and faculty members of the computing department of one private University in Metro Manila, Philippines, had the

443

chance to experience full online learning. During that period, much data can already be obtained to understand the possible challenges of students in an online learning environment.

This study was conducted using descriptive statistics whose data was made available through a survey utilizing MS Forms Pro. Before the conduct of the survey, the researchers asked for some inputs from selected students about their difficulties in online learning. This served as the basis for the revised questionnaire. The survey was administered by the student council of the computing department to avoid possible biases and influence from the faculty members.

The online survey was conducted from March 16 until March 18, 2020, where most students were still online. For the three (3) day period, a total of 300 responses were obtained. The survey questionnaire includes questions attributed to 1.) Availability of Devices, 2.) Internet Connectivity and Reliability, and 3.) Other issues that may hamper the students' online engagement. Aside from the multiple-choice option, an open-ended question was included to give them the chance to express their thoughts on how the department can improve their learning experience online.

## RESULTS AND DISCUSSION

Below presents the results of the online survey conducted among the students of one private University in Manila. Table 1 shows the breakdown of the respondents.

Table 1. Demographics

| Categories | Total Number | Frequency |
|---|---|---|
| Degree Courses | 58 | 19% |
|    BSCS | 242 | 81% |
|    BSIT | | |
| Year Level | | |
|    1$^{st}$ Year | 105 | 35% |
|    2$^{nd}$ Year | 140 | 47% |
|    3$^{rd}$ year / 4$^{th}$ year | 55 | 18% |
| Total | 300 | 100% |

From the 300 respondents, 58 were BSCS students, and 242 were from the BSIT program. This number was not as surprising as most enrollees in the computing department were enrolled in the latter. For the year level, the 2nd year or sophomore students had the greatest number of responses yielding 140 or 47% of the entire population. This is followed by 1st year or freshmen students with total respondents of 104 representing 35% of the whole respondents and, lastly, the irregular students (3$^{rd}$-year or 4th-year students) comprising only 55 or 18% of the entire respondents. The Philippines implemented the K to 12 Basic Education Program, and there are still no regular students

444

in their third or fourth year of tertiary education. The small number of respondents from the 3rd-year and 4th-year students was those who did not go through the program.

For the year level, the 2nd year or sophomore students had the greatest number of responses yielding 140 or 47% of the entire population. This is followed by 1st year or freshmen students with total respondents of 104 representing 35% of the whole respondents and, lastly, the irregular students (3rd Year or 4th-year students) comprising only 55 or 18% of the entire respondents. The Philippines implemented the K to 12 Basic Education Program, and there are still no regular students in their third or fourth year of tertiary education. The small number of respondents from the 3rd year and 4th year students were those who did not go through the program.

## Devices used for Online Learning

The next two Figures (1 and 2) represent the devices used by the students in online learning. Figure 1 shows the result per program, whereas Figure 2 describes the result per year level

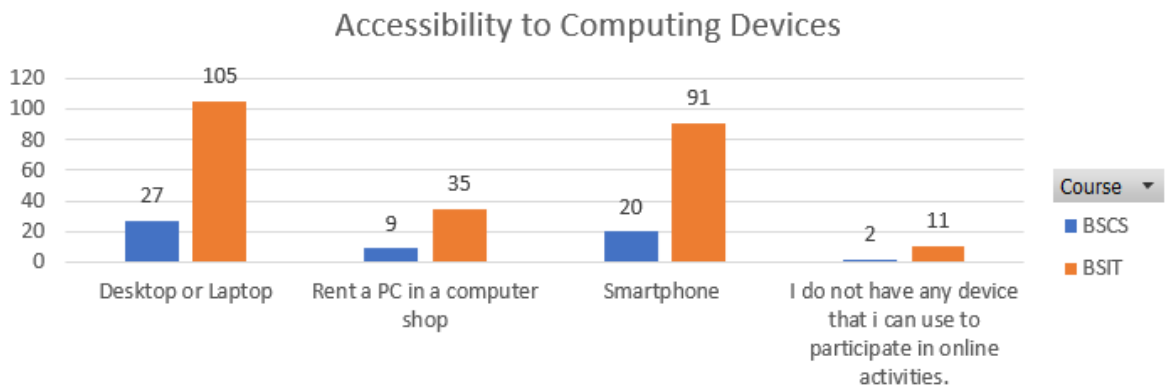

*Figure 1.* Devices used in Online Learning per Program

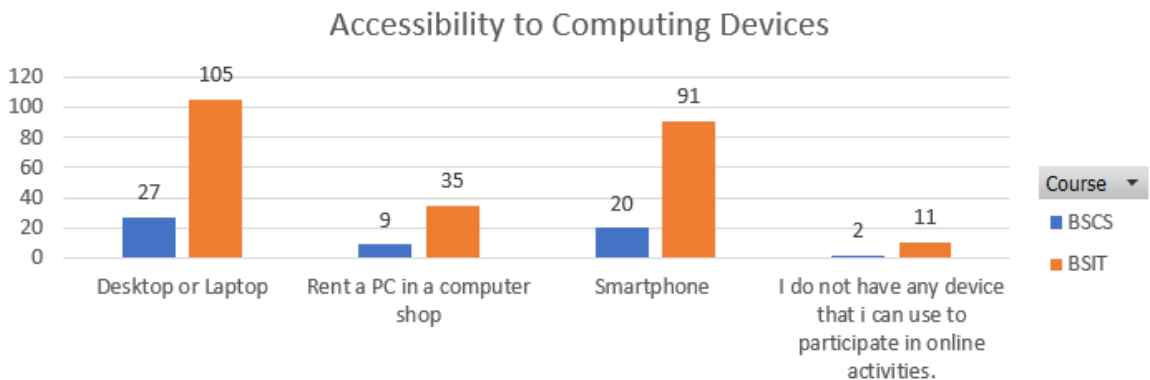

*Figure 2.* Devices used in Online Learning per Year Level

Figure 1 shows that almost half of the respondents are using computers and laptops for online learning. Out of the 234 respondents from the IT program, 105 or 44.87 % use a desktop computer or laptop, while 35 or 14.95% of students have to rent a PC in a computer or internet shop to participate in online learning. On the other hand, 91 or 38.8% of them are using smartphones to access the Internet. Only 11 or 0.04% do not have any device that can be used for online learning. The same pattern is also true for CS students. From a total of 58 respondents, 50% have computers or laptops, while 9 or 15.51% rent a computer, and 20 or 34% make use of smartphones. Two (2) or 3% of CS students do not have access to the Internet. This outlook represents a big issue for computing students doing online learning. As already mentioned, specific devices should be available as some professional courses would require minimum device specifications. And one of these is a laptop or a desktop PC ("Student Computing Requirements - Information Technology - University of Florida," n.d.).

In terms of how it varies per year level, looking at Figure 2, we can infer that most of the 2nd year students make use of desktop PCs or laptops to participate in online learning in comparison to freshmen students. Comparing to 55.71% (78 out of 140) of sophomore students, freshmen students only have 29.5% (31 out of 105) who make use of PCs or laptops. However, if combined with the computer rental, that would make 49.5% (52 out of 105) of the entire freshmen students. In terms of using smartphone devices, it can be observed that 28.5% (50 out of 140) of IT students and 47.61% (50 out of 105) of CS students can engage in online learning through their mobile devices. With the advancement in mobile technologies, modern smartphones can now perform some flexibilities of a computer (Fabito, 2017). Additionally, some Windows and Mac-based applications can now be installed in Android and iOS-based smartphones making online-based applications for PCs available in smartphones (Pilar, Jorge, & Cristina, 2013). Hence, some students rely on smartphones if PCs and laptops are not available.

*Internet Accessibility*

While internet accessibility is one of the crucial components for a student to participate in online learning, Figure 3 presents the variation of how students access the Internet. From the chart, it can be noted that very few have no access to the Internet. This only represents 8% (24 out of 300) of the entire population. The twenty-four responses are a combination of those who do not have an appropriate device (Figure 1) and a reliable internet connection (Figure 5)

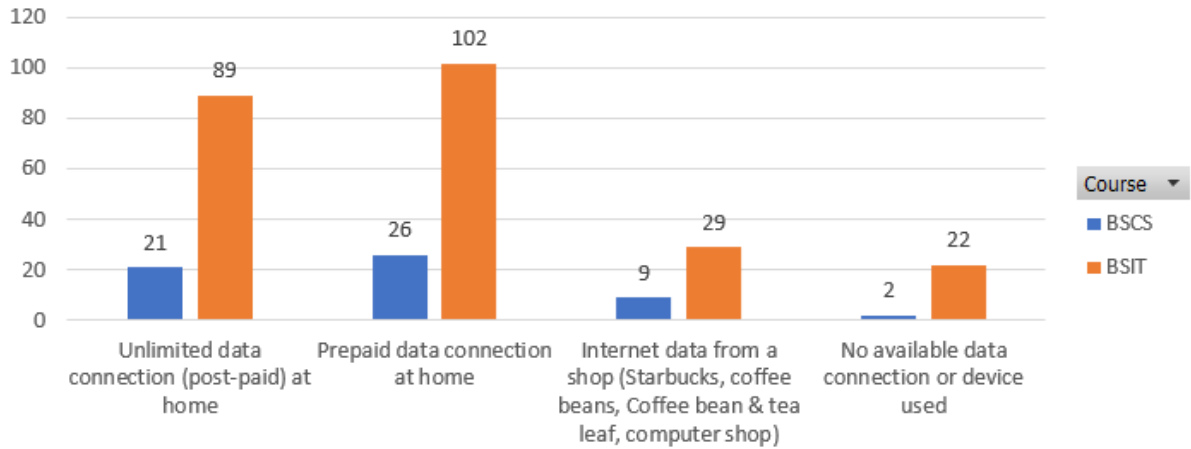

*Figure 3.* Internet Accessibility of Students

Merging the first two items (Availability of devices and Internet accessibility), Figure 4 illustrates the data connectivity available for the devices used by the students. For desktop or laptop users, out of the 132 users, 82 or 62% have an unlimited data connection. This is followed by 48 or 36% on prepaid data connection, and the remaining two (2) or 2 gets the internet connection from a computer shop or has no available connection. As already stated, while a personal computer or laptop is necessary for computing courses, an Internet connection is also essential to participate in live meetings and online assessments. Looking at the data, only 27% (82 out of 300) of the entire respondents may have the full potential to take advantage of online learning - a computer and an unlimited internet connection. For those using smartphones, out of the 111 users, it can be observed that the majority (72 or 65%) of the smartphone users are connected to a prepaid internet service connection. In contrast, only 25 or 23% have an unlimited data connection. This goes to show that smartphone users tend to spend more just to be connected to the Internet. According to a news article, the Philippines, aside from having the slowest Internet connection in Asia, it is also one of the most expensive ("LIST: Philippines ranks 21st of 22 Asian countries in Internet download speed," n.d.) Overall, the data tells us that most PC and Laptop users have an unlimited internet connection, while smartphone users tend to use prepaid data for their online needs.

From the 44 students who answered that they rent a PC to a computer shop, only 29 or 66 % responded that they get an Internet connection from the shop. The other 8 or 18% do not have an Internet connection. This may be true since some computer shops only offer network games. The computers utilized in this case are for solving offline computing problems (e.g., programming activities). The last 7 or 16% may seem irrelevant as it cannot be justified conclusively.

The authors, however, agree that there might be a few items that have not been adequately answered by the respondents. The same goes for the 13 respondents who noted that they do not have any device that can be used for online learning. From the 13 responses, only 8 or 62% have recorded that they do not have an internet connection or do not have appropriate devices. The remaining 5 or 38% responses seemed to be illogical. However, given some minute discrepancies, the general picture observed in both surveys tells us that more than half of the respondents do not have an adequate device to perform computing activities meant to be answered outside the school campus. Additionally, a continuous Internet connection also holds a challenge to everyone as it is limited only to a few students. The reliability of the Internet connection, however, is a different story that will be discussed in the next section.

*Internet Reliability*

As already stated, a continuous internet connection is necessary for students to fully engaged in online learning. This is true, especially when an online meeting is required for all subjects enrolled. However, variables may come to play that would inhibit continuous internet connection. Aside from being one of the most expensive internet providers in Asia, the Philippines is also one of the slowest ("LIST: Philippines ranks 21st of 22 Asian countries in Internet download speed," n.d.). This can be reflected in the responses of the students. As observed, out of the 300 students, only thirty-eight (38) or 13%0 of them are experiencing a fast and reliable internet connection.

On the other hand, sixty (60) or 20% are experiencing relatively fast internet connection but is not always available. It often happens as there are times when the internet connection is strong and would then subside in some occasions. In terms of the poor connection, one-hundred thirteen (113) or 37% is experiencing slow connection but sufficient to meet the requirement for their online activities. Overall, it can be interpreted that two hundred eleven (211) or 70% of the respondents can somehow meet their internet requirements for online learning.

Only seventy-seven (77) or 26% are experiencing slow connection and are not able to meet their online requirements. This is followed by the twelve (12) or 4% without an internet connection used at home.

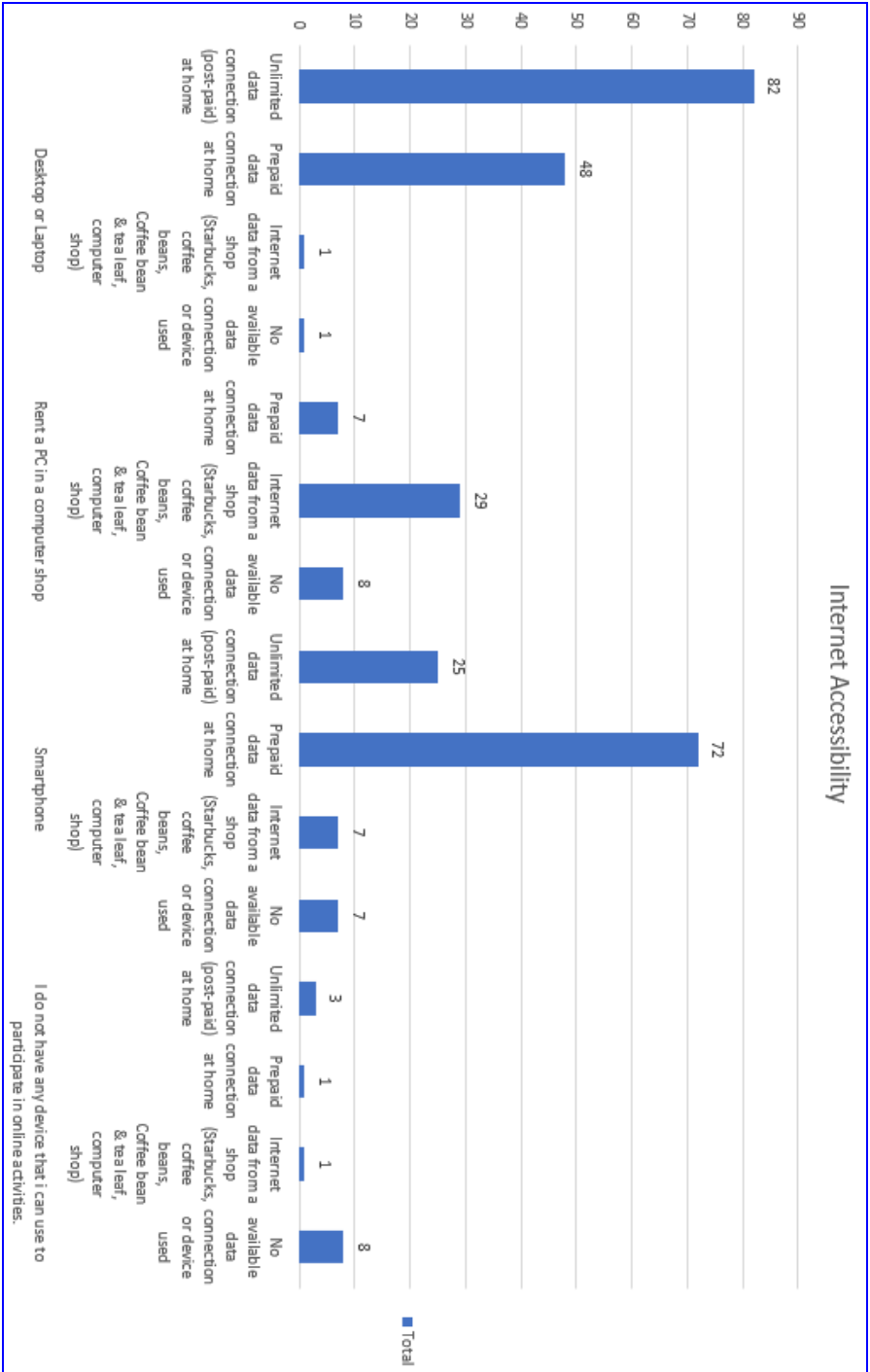

*Figure 4.* Internet Accessibility on the devices used.

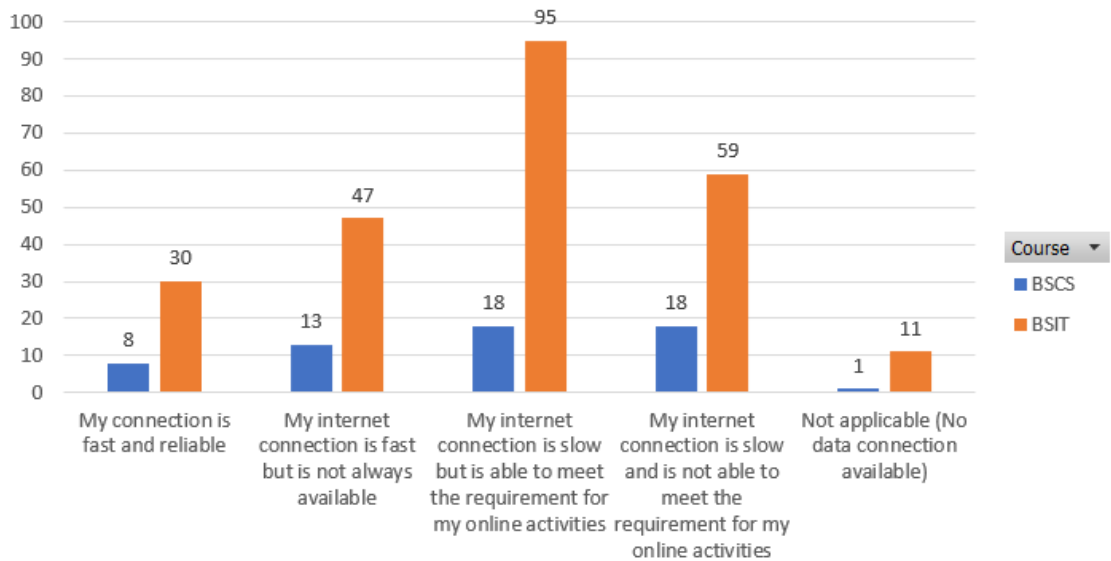

*Figure 5.* Internet Reliability

Translating the reliability of the internet connection on the devices used by the students, Figure 7 shows the comparison. The thirty-eight (38) responses representing the fast and reliable connection in Figure 5 are shared among desktop or laptop users (30), smartphone users (7), and computer shop users (1). From the 132 desktop or laptop users, 30 or 23% have a fast and reliable internet connection. The other 31 or 23% is experiencing fast connection but is not always reliable. What is obvious in the figure is the fact that the majority of the respondents are experiencing slow connection but can fulfill their online needs. This corresponds to 51 or 39% of the entire responses. Only 19 or 14% of the 132 users have difficulty complying with the online learning requirements due to the slow internet connection.

This trend is also the same for smartphone users. Out of the 111 respondents, 7 or 6% experienced fast and reliable internet connection, 22 or 20% experienced fast but not always a reliable connection, 49 or 44% experienced slow connection but is relatively able to meet their online requirements, 32 or 29% experienced slow connection and is not able to meet their online requirements and 1 or 1% without a data connection.

For those who use computer shops, it is interesting to find out that out of the 44 users, 54% of the users (24 responses) noted that the internet connection found on computer or coffee shops are slow and is not able to meet their online learning internet connectivity reliability. Only 16 or 36% can be identified as having a stable connection to meet internet demand. Looking at the figures, another problem that computers or coffee shop users tend to experience is the availability of internet reliability. On the other hand, from the 13 responses that noted that they do not have any devices used for online learning, 6 or 46% answered that they do not have an internet connection. The remaining 8 or 54% seemed to have an internet connection but does not have any device. This, however, still needs validation, which the study was not able to perform.

Aside from the availability of devices, internet connectivity, and reliability, the survey also looked at other issues that may hamper the students' online learning experience. The questionnaire was partially obtained from students whom the authors had interviewed before the survey was done. The items include internet accessibility and reliability (as a way to validate the first three surveys), study areas, and the learning itself. Respondents were able to select any of the choices portrayed in Figure 6.

Overall, the top three (3) issues selected, highlighted in color green, were the following: 1.) The difficulty of clarifying topics or discussions with the professor, which accounts for 169 or 56% of the 300 responses. 2.) Poor internet connection which accounts for 168 or 56%, and 3.) Lack of good study area for online learning. This accounts for 144 or 48% of the respondents.

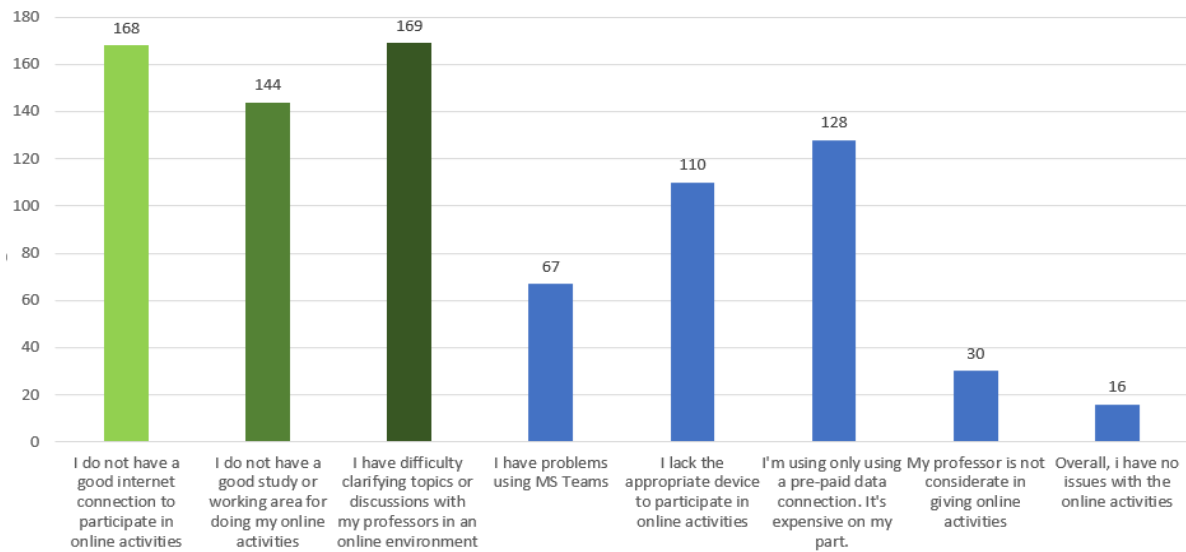

*Figure* 6. General Issues and concerns

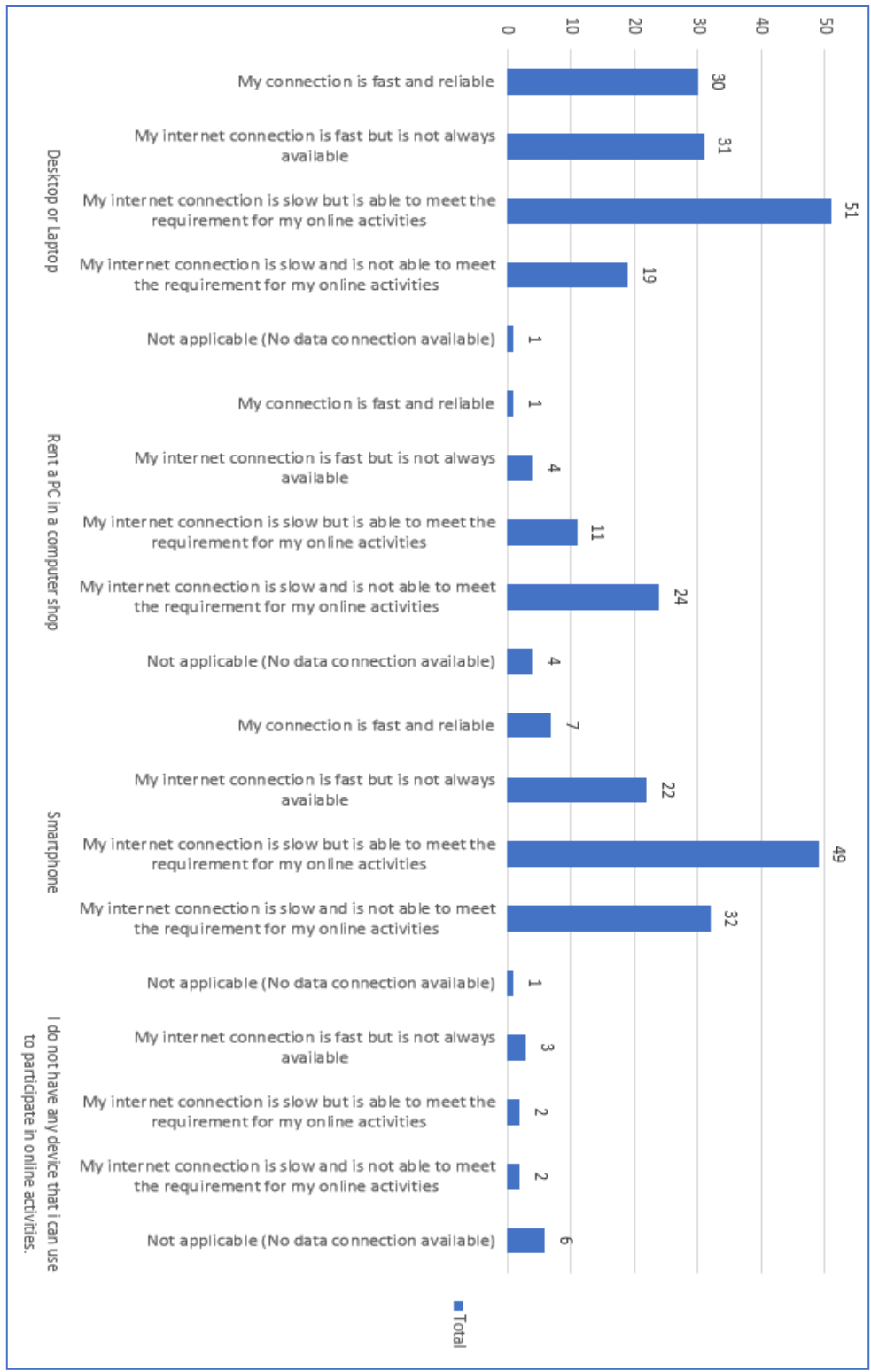

*Figure 7.* Internet Reliability on the devices used

From the result, it seemed that it is not just the devices or the internet connectivity that is the problem addressed by the students in an online learning environment but also the learning itself. Since online learning is limited to providing face-to-face engagement and interactive discussion, a pedagogical design for online education should be well established ("Five Pedagogical Practices to Improve Your Online Course," n.d.). Faculty members need to learn certain skills and carefully craft methodologies suitable for online learning, which includes the design, delivery, and assessment while providing social, affective, and emotional support to students (Sharma, 2018). Only through a well designed blended or online learning pedagogy would the learners and teachers fully obtain the advantage of online learning (Heggart & Yoo, 2018; Shorey et al., 2018). Admittedly, when faculty members were advised to do full online learning due to the Enhanced Community Quarantine (ECQ) in Luzon, faculty members were not trained on how to do online learning effectively. In recent years, the LMS used by the computing department was used primarily as a repository for class materials, uploading of assignments, and for online exams. It was the first time for the computing department to make full-blown online learning. Hence, there is a possibility of failing to adapt to the learners' needs in online learning. This can be further validated by the open-ended question included in the survey. Answers include that some faculty members from other departments were merely adding videos and other learning resource materials in the Learning Management System (LMS) used by the computing department. This is one problem that may be common among faculty members if they are not adequately trained to shift to online learning.

The second (2nd) barrier observed from the survey is the internet connectivity. This is a validation found in Figure 3, which shows that very few are connected to a fast and reliable internet connection. Lastly, the third (3rd) issue observed is the lack of a functional study area for the learners. Unlike in a school setting, where learning is more conducive, students were not adequately prepared to go online learning at home. The abrupt decision to go online due to the ECQ brought about by the COVID-19 pandemic has become an eye-opener in terms of the lack of preparedness of students and faculty members to go online. Although an LMS is employed in the computing department for years now, it goes to show that both students and faculty members are not prepared to do full online learning for an extended period.

Surprisingly, the survey has also revealed that only one hundred ten (110) selected "I lack the device to participate in online activities", which is the 5th among the eight issues provided. Although there were only one hundred thirty-two (132) who selected that they have laptops or PCs, it was not the priority in terms of their concerns. As already mentioned, PCs and Laptops are necessary when performing programming or other computing problems. While several reasons can explain this, one conclusive justification could be that since the majority of the students are just starting in their career (1st year and 2nd Year), most of their subjects are not professional subjects but minor subjects. Another possible reason is that faculty members may be cautious about providing computing problems as they are aware of the difficulties faced by the students. However, a follow-up study must still be conducted to understand the result further.

## CONCLUSIONS AND RECOMMENDATIONS

This study has provided insights concerning the barriers of computing students in online learning. Looking at the results of the survey, it shows that aside from the lack of computers or laptops, which is necessary for solving computing problems, internet connectivity also poses a problem. To aid the need for a desktop PC or laptops for students, the development of mobile apps may be explored that can help simulate professional subjects such as programming and web development. Mobile apps may provide functionalities that are observable in computers. However, the effectiveness of the app is another endeavor that researchers can investigate.

The survey has also shown that faculty members may have failed to adapt to the needs of the students in an online learning environment. Hence, there is a need for faculty members to undergo training that would allow them to design a pedagogy suited for online learning. The management should adopt a policy that would institute continuous training and monitoring for faculty members to address the pedagogical concerns of students in online learning.

While the primary data of the study mainly came from the students, it would also be an excellent addition to understand the perspective of the faculty members in terms of their experiences with their students. Their insights could help validate the responses in the survey and provide other barriers that may not have been included in the study. Subsequently, studying the restrictions of faculty members in online learning, including their participation, would also be necessary when crafting a holistic framework encompassing both the learners and teachers for successful online learning for the computing students. Once a framework is developed and tested, similar to the study of (Han, Wang, & Jiang, 2019), it can help provide administrators and policymakers advance online learning implementation in the University.

## ACKNOWLEDGEMENT

The authors acknowledge the Center for Research, National University-Manila, for their support in the conduct of the study.

# Appendix A. Questionnaire

# Questionnaire

This survey is intended to understand the practices used by ____**STUDENTS** in light of the implementation of Blended Learning for the entire duration of the suspension of classes due to the Enhanced Community Quarantine (ECQ) brought about by COVID-19. Data collected in this survey will be used solely to help address issues related to blended learning. Your name and section will not be collected

1. Course
   BSCS
   BSIT

2. Year Level
   $1^{st}$ Year
   $2^{nd}$ Year
   $3^{rd}$ year or $4^{th}$ Year

3. Which device are you using to participate in your online activities
   Desktop or Laptop

   Smartphone

   Rent a PC in a computer shop

   I do not have any device that I can use to participate in online activities.

4. What type of data connection are you using for your device
   Unlimited data connection (postpaid) at home

   Prepaid data connection at home

   Internet data from a shop (Starbucks, coffee beans, Coffee bean & tea leaf, computer shop)

   No available data connection or device used

5. How would you best describe your device in terms of satisfying the requirements for your online activities (e.g., Online meeting, online exam, and other online activities)
   My device can perform all of the functionalities required for my online activities

   My device can perform most of the functionalities required for my online activities

   My device can perform only a few of the functionalities required for my online activities

   My device is not able to perform any functionalities required for my online activities

   Not Applicable (No device)

6. How would you best describe your data connectivity in terms of speed and reliability
   My connection is fast and reliable

   My internet connection is fast but is not always available

   My internet connection is slow but can meet the requirement for my online activities

   My internet connection is slow and is not able to meet the requirement for my online activities

   Not applicable (No data connection available)

7. How would you describe your learning experience in the blended learning environment
   I can follow and catch-up with the lessons and activities provided in the online environment.

I can sometimes follow, and catch-up with the lesson's activities provided in the online activities

I am not able to follow and catch-up with the lessons and activities provided in the online activities

8. How would you best describe your professors in terms of delivery of instruction in the online environment (Major Subjects)

My professors can provide adequate resources and time to meet us (online) and give necessary feedback to our activities.

Some of my professors can provide adequate resources and time to meet us (online) and give necessary feedback to our activities.

My professors are not able to provide adequate resources and time to meet us (online) and give necessary feedback to our activities.

9. How would you best describe your professors in terms of delivery of instruction in the online environment (Minor Subjects)

My professors can provide adequate resources and time to meet us (online) and give necessary feedback to our activities.

Some of my professors can provide adequate resources and time to meet us (online) and give necessary feedback to our activities.

My professors are not able to provide adequate resources and time to meet us (online) and give necessary feedback to our activities.

General Issues

I have problems using MS Teams

I lack the appropriate device to participate in online activities

I do not have a good internet connection to participate in online activities

I'm using only using a prepaid data connection. It's expensive on my part.

My professor is not considerate in giving online activities

I have difficulty clarifying topics or discussions with my professors in an online environment

I do not have a good study or working area for doing my online activities

Overall, I have no issues with the online activities